\documentclass[prd,aps,floatfix,nofootinbib,11 pt]{revtex4}
\usepackage{amsmath,graphicx,color,epsfig}

\input{tcilatex}

\begin{document}

\title{Effect of flipping noise on the entanglement dynamics of a $3$x$3$
system}
\author{M. Ramzan\thanks{%
mramzan@phys.qau.edu.pk}}
\address{Department of Physics Quaid-i-Azam University \\
Islamabad 45320, Pakistan}
\date{\today }

\begin{abstract}
Entanglement dynamics of a qutrit-qutrit system under the influence of
global, local and multilocal decoherence introduced by phase flip, trit flip
and trit phase flip channels is investigated. The negativity and realignment
criterion are used to quantify the entanglement of the system. It is shown
that the entanglement sudden death and distibility sudden death can be
avoided in the presence of phase flip, trit flip and trit-phase flip
environments. It is shown that certain free entangled distillable qutrit
states become bound entangled or separable i.e. convert into non-distillable
states under different flipping noises. It is also seen that local
operations do not have any effect on the entanglement dynamics of the
system. Further more, no ESD and DSD is seen for the case of trit flip
channel.\newline
\end{abstract}

\maketitle
\date{\today}

\address{Department of Physics Quaid-i-Azam University \\
Islamabad 45320, Pakistan}


\vspace*{1.0cm}

\vspace*{1.0cm}



\section{Introduction}

Quantum entanglement is a fundamental resource for many quantum information
processing tasks, e.g. super-dense coding, quantum cryptography and quantum
error correction [1-4]. Entangled states can be used in constructing number
of protocols, e.g. teleportation [5], key distribution and quantum
computation [6]. During recent past, entanglement sudden death (ESD) has
been investigated by different authors for bipartite and multipartite states
[7-10]. Yu and Eberly [11, 12] have shown that entanglement loss occurs in a
finite time under the action of pure vacuum noise in a bipartite qubit
system. A geometric interpretation of the phenomenon of ESD has been given
in Ref. [13]. Furthermore, experimental evidences of ESD have been reported
for optical setups [14, 15] and atomic ensembles [16]. Peres-Horodecki [17,
18] have studied entanglement of qubit-qubit and qubit-qutrit states and
established separability criterion. According to this criterion, the partial
transpose of a separable density matrix must have non-negative eigenvalues.
For non-separable states, the sum of the absolute values of the negative
eigenvalues of the partial transpose matrix gives the degree of entanglement
of a density matrix also termed as negativity. Ann et al. [19] have studied
a qubit-qutrit system where they have shown the existence of ESD under the
influence of dephasing noise.

Bipartite entangled states are divided into free-entangled states and
bound-entangled states [20, 21]. Free-entangled states can be distilled
under local operations and classical communication (LOCC) whereas
bound-entangled states can not be distilled to pure-state entanglement
irrespective of the number of copies of the initial state available. Wang et
al. proposed that free-entangled states may be converted into
bound-entangled states under multi-local [22] and collective dephasing
processes [23]. They have shown that certain free-entangled states of
qutrit-qutrit systems become non-distillable in a finite time under the
influence of classical noise. For a qutrit-qutrit system there are many
bound entangled states and no single criterion can fully describe all of
them [24]. However, the realignment criterion can be used to detect certain
bound entangled states [25].

Since, it is\ not possible to completely isolate a quantum system from its
environment. Therefore, one needs to investigate the behavior of
entanglement in the presence of environmental effects. A major problem of
quantum communication is to faithfully transmit unknown quantum states
through a noisy quantum channel. When quantum information is sent through a
channel, the carriers of the information interact with the channel and get
entangled with its many degrees of freedom. In order to quantify
entanglement, two measures are the negativity [26], a measure of a state
having negative partial transpose and the realignment criterion [25]. The
negativity is equal to the absolute value of the sum of negative eigenvalues
of partial transpose of a state. When the negativity becomes zero, we need
to study the time evolution of realignment criterion to determine the
existence of bound-entangled states. However, it should be kept in mind that
realignment criterion could not detect all entangled states.

In this paper, the effect of flipping noise on the entanglement dynamics of
a qutrit-qutrit system under the influence of global, local and multilocal
decoherence is investigated by considering phase flip, trit flip and trit
phase flip channels. It is shown that the entanglement sudden death can be
avoided in the presence of phase flip, trit flip and trit phase flip
environments. It is also shown that particular local operations cannot avoid
the non-distillability of the distillable states. Further more, no ESD and
DSD is seen for trit flip channel in the presence of global noise.

\section{Dynamics of a qutrit-qutrit system under flipping noise}

Let us consider that the initial state is interacting with the noisy
environment both collectively and individually. Local and multi-local
couplings describe the situation when both the qutrits are independently
influenced by their individual noisy environments. Whereas, the global
decoherence corresponds to the situation when it is influenced by both local
and multilocal noises at the same time. The state shared by the two parties
is an entangled qutrit-qutrit state of the form%
\begin{equation}
\rho (0)=\frac{2}{7}\left. \left( (|\Psi _{+}\rangle \left\langle \Psi
_{+}\right\vert +\frac{\alpha }{7}\sigma _{+}+\frac{5-\alpha }{7}\sigma
_{-}\right) \right.
\end{equation}%
where $2\leq \alpha \leq 5$, $|\Psi _{+}\rangle =\frac{1}{\sqrt{3}}%
(|00\rangle +|11\rangle +|22\rangle )$ is a maximally entangled bipartite
qutrit state and $\sigma \pm $ are separable states given as $\sigma
_{+}=1/3(|01\rangle \left\langle 01\right\vert +|12\rangle \left\langle
12\right\vert +|20\rangle \left\langle 20\right\vert )$, $\sigma
_{-}=1/3(|01\rangle \left\langle 01\right\vert +|21\rangle \left\langle
21\right\vert +|02\rangle \left\langle 02\right\vert )$

The above density matrix is separable for $2\leq \alpha \leq 3$, bound
entangled for $3\leq \alpha \leq 4$ and free entangled for $4<\alpha \leq 5$
[27]. The interaction between the system and its environment introduces the
decoherence to the system, which is a process of the undesired correlation
between the system and the environment. The dynamics of the composite system
in the presence of flipping noise can be described in Kraus operators
formalism [29]

\begin{equation}
\rho _{f}=\sum_{k}E_{k}\rho _{i}E_{k}^{\dag },  \label{E5}
\end{equation}%
where the Kraus operators $E_{i}$ satisfy the following completeness relation

\begin{equation}
\sum_{k}E_{k}^{\dag }E_{k}=I.  \label{5}
\end{equation}
The single qutrit Kraus operators for phase flip channel can be written as

\begin{equation}
E_{0}=\sqrt{1-\frac{2p}{3}}\left(
\begin{array}{ccc}
1 & 0 & 0 \\
0 & 1 & 0 \\
0 & 0 & 1%
\end{array}%
\right) ,\ \ E_{1}=\sqrt{\frac{p}{3}}\left(
\begin{array}{ccc}
1 & 0 & 0 \\
0 & e^{\frac{-2\pi i}{3}} & 0 \\
0 & 0 & e^{\frac{2\pi i}{3}}%
\end{array}%
\right) ,\ \ E_{2}=\sqrt{\frac{p}{3}}\left(
\begin{array}{ccc}
1 & 0 & 0 \\
0 & e^{\frac{2\pi i}{3}} & 0 \\
0 & 0 & e^{\frac{-2\pi i}{3}}%
\end{array}%
\right) ,  \label{E7}
\end{equation}%
and the single qutrit Kraus operators for the trit flip channel are given as

\begin{equation}
E_{0}=\sqrt{1-\frac{2p}{3}}\left(
\begin{array}{ccc}
1 & 0 & 0 \\
0 & 1 & 0 \\
0 & 0 & 1%
\end{array}%
\right) ,\ \ E_{1}=\sqrt{\frac{p}{3}}\left(
\begin{array}{ccc}
0 & 0 & 1 \\
1 & 0 & 0 \\
0 & 1 & 0%
\end{array}%
\right) ,\ \ E_{2}=\sqrt{\frac{p}{3}}\left(
\begin{array}{ccc}
0 & 1 & 0 \\
0 & 0 & 1 \\
1 & 0 & 0%
\end{array}%
\right)  \label{7}
\end{equation}%
Whereas, the single qutrit Kraus operators for the trit phase flip channel
are given by

\begin{eqnarray*}
E_{0} &=&\sqrt{1-\frac{2p}{3}}\left(
\begin{array}{ccc}
1 & 0 & 0 \\
0 & 1 & 0 \\
0 & 0 & 1%
\end{array}%
\right) ,\ \ E_{1}=\sqrt{\frac{p}{3}}\left(
\begin{array}{ccc}
0 & 0 & e^{\frac{2\pi i}{3}} \\
1 & 0 & 0 \\
0 & e^{\frac{-2\pi i}{3}} & 0%
\end{array}%
\right) , \\
E_{2} &=&\sqrt{\frac{p}{3}}\left(
\begin{array}{ccc}
0 & e^{\frac{-2\pi i}{3}} & 0 \\
0 & 0 & e^{\frac{2\pi i}{3}} \\
1 & 0 & 0%
\end{array}%
\right) ,\ \ E_{3}=\sqrt{\frac{p}{3}}\left(
\begin{array}{ccc}
0 & e^{\frac{2\pi i}{3}} & 0 \\
0 & 0 & e^{\frac{-2\pi i}{3}} \\
1 & 0 & 0%
\end{array}%
\right)
\end{eqnarray*}%
where $p=1-e^{-\Gamma t}$ represents the quantum noise parameter usually
termed as decoherence parameter. Here the bounds [0, 1] of $p$ correspond to
$t=0$, $\infty $ respectively. The evolution of the initial density matrix
of the composite system when it is influenced by local and multi-local
environments is given in Kraus operator form as
\begin{equation}
\rho _{f}=\sum\limits_{i,j,k}(E_{j}^{B}E_{k}^{A})\rho
_{AR}(E_{j}^{B}E_{k}^{A})^{\dagger }
\end{equation}%
and the evolution of the system when it is influenced by global environment
is given in Kraus operator representation as
\begin{equation}
\rho _{f}=\sum\limits_{i,j,k}(E_{i}^{AB}E_{j}^{B}E_{k}^{A})\rho
_{AR}(E_{i}^{AB}E_{j}^{B}E_{k}^{A})^{\dagger }
\end{equation}%
where $E_{k}^{A}=E_{m}^{A}\otimes I_{3},$ $I_{2}\otimes E_{j}^{B}$ are the
Kraus operators of the multilocal couplings of the individual qutrits and $%
E_{i}^{AB}=E_{m}^{A}\otimes E_{n}^{A}$ are the Kraus operators of the
collective coupling of the qutrit system. Using equations (2-5) along with
the initial density matrix of as given in equation (1) and taking the
partial transpose over the second qutrit, the eigenvalues of the final
density matrix can be easily found. Let the decoherence parameters for local
and global noise be $p_{1}$, $p_{2}$ and $p$ respectively. The eigenvalues
of the partial transpose matrix, when both qutrits are influenced by
multi-local noise of the phase flip channel, are given as
\begin{eqnarray}
\lambda _{1} &=&\lambda _{2}=\lambda _{3}=\frac{2}{21}  \notag \\
\lambda _{4,5,6} &=&\frac{1}{42}[5-\left. \sqrt{%
\begin{array}{c}
4\text{$\alpha $}^{2}-20\text{$\alpha $}+16\text{$p_{1}$}^{2}(\text{$p_{2}$}%
-1)^{2} \\
-32\text{$p_{1}$}(\text{$p_{2}$}-1)^{2}+16\text{$p_{2}$}^{2}-32\text{$p_{2}$}%
+41%
\end{array}%
}\right. ] \\
\lambda _{7,8,9} &=&\frac{1}{42}[5+\left. \sqrt{%
\begin{array}{c}
4\text{$\alpha $}^{2}-20\text{$\alpha $}+16\text{$p_{1}$}^{2}(\text{$p_{2}$}%
-1)^{2} \\
-32\text{$p_{1}$}(\text{$p_{2}$}-1)^{2}+16\text{$p_{2}$}^{2}-32\text{$p_{2}$}%
+41%
\end{array}%
}\right. ]
\end{eqnarray}%
and the realignment criterion in this case reads%
\begin{equation}
||\rho ^{R}||-1=\frac{2}{21}\sqrt{3\text{$A_{1}$}^{2}-15\text{$A_{1}$}+19}+%
\frac{4}{7}\sqrt{(\text{$p_{2}p_{1}$}-\text{$p_{1}$}-\text{$p_{2}$}+1)^{2}}-%
\frac{26}{3}
\end{equation}%
The expressions for local noise can be obtained by setting $p_{2}=0$. The
eigenvalues of the partial transpose matrix, when both the qutrits are
influenced by the global noise of phase flip channel, are given by%
\begin{eqnarray}
\lambda _{1} &=&\lambda _{2}=\lambda _{3}=\frac{2}{21} \\
\lambda _{4,5,6} &=&\frac{{1}}{42}\left( \sqrt{%
\begin{array}{c}
16p^{8}-128p^{7}+448p^{6}-896p^{5}+1120p^{4} \\
-896p^{3}+448p^{2}-128p+4\text{$\alpha $}^{2}-20\text{$\alpha $}+41%
\end{array}%
}+5\right) \\
\lambda _{7,8,9} &=&\frac{{1}}{42}\left( \sqrt{%
\begin{array}{c}
16p^{8}-128p^{7}+448p^{6}-896p^{5}+1120p^{4} \\
-896p^{3}+448p^{2}-128p+4\text{$\alpha $}^{2}-20\text{$\alpha $}+41%
\end{array}%
}-5\right)
\end{eqnarray}%
and the realignment criterion in this case reads%
\begin{equation}
||\rho ^{R}||-1=\frac{4p^{4}}{7}-\frac{16p^{3}}{7}+\frac{24p^{2}}{7}-\frac{%
16p}{7}+\frac{2}{21}\sqrt{3\text{$\alpha $}^{2}-15\text{$\alpha $}+19}-\frac{%
2}{21}
\end{equation}%
The eigenvalues of the partial transpose matrix when both the qutrits are
influenced by the trit flip and trit phase flip channels are not provided
with their analytical relations as these are too lengthy in size. Therefore,
they are just interpreted in the graphs. The entanglement for all mixed
states $\rho _{AB}$ can be quantified by the negativity%
\begin{equation}
N(\rho _{AB})=\max \{0,\sum\limits_{k}\left\vert \lambda
_{k}^{T_{A}(-)}\right\vert \})
\end{equation}%
where $\lambda _{k}^{T_{A}(-)}$\ represents the negative eigenvalues of the
partial transpose of the density matrix $\rho _{AB}$.

\section{Discussions}

In this work, the effect of decoherence on a qutrit-qutrit system is
investigated. In figure 1, the negativity and realignment criterion are
plotted as a function of decoherence parameter $p$ for local noise
introduced through phase flip, trit flip and trit phase flip channels. It is
seen that the partial transpose criterion fails to detect the behavior of
entanglement in density matrices of larger $\alpha $ for higher values of
decoherence. The quantity $||\rho ^{R}||-1$ is positive for $p<0.5$ which
shows that these density matrices are bound entangled. However for $p>0.5$,
the realignment criterion also fails to detect the possible entanglement for
these states.

In figure 2, negativity and realignment criterion are plotted as a function
of decoherence parameter $p$ for multi-local noise introduced through phase
flip, trit flip and trit phase flip channels. It can be seen that no ESD
occurs in any density matrix as the negativity is positive for all the
density matrices. However, for $p<0.1$, some states that correspond to large
values of $\alpha $ become positive partial transpose states (PPT).

In figure 3, negativity and realignment criterion are plotted as a function
of decoherence parameter $p$ for global noise introduced through phase flip,
trit flip and trit phase flip channels. It is seen that the free entangled
distillable states convert into bound entangled or separable states and
therefore become completely non-distillable in the presence of global noise.
In figure 4, three-dimensional graphs for negativity are plotted as a
function of decoherence parameter $p$ and parameter $\alpha $ for
multi-local and global noises introduced through phase flip, trit flip and
trit phase flip channels. It is seen that ESD and DSD can be completely
avoided for the case of trit flip channel.

In order to see the effect of local unitary operation on the dynamics of
qutrit-qutrit state, let the unitary operator $U$ = $I_{3}\otimes \theta $,
with $\theta =|0\rangle \left\langle 1\right\vert +|1\rangle \left\langle
0\right\vert +|2\rangle \left\langle 2\right\vert $. By applying this
operator locally to the state (equation 1), it converts the maximally
entangled state $|\Psi _{+}\rangle $ into another maximally entangled state
given by $|\tilde{\Psi}_{+}\rangle =\frac{1}{\sqrt{3}}(|01\rangle
+|10\rangle +|22\rangle )$ and the two separable states $\sigma _{\pm }$ are
transformed to $\tilde{\sigma}_{+}=1/3(|00\rangle \left\langle 00\right\vert
+|12\rangle \left\langle 12\right\vert +|21\rangle \left\langle
21\right\vert )$, $\sigma _{-}=1/3(|11\rangle \left\langle 11\right\vert
+|20\rangle \left\langle 20\right\vert +|02\rangle \left\langle
02\right\vert ),$ the final density matrix takes the form
\begin{equation}
\sigma _{\alpha }=U\rho _{\alpha }U^{\dagger }=\frac{2}{7}\left. \left( (|%
\tilde{\Psi}_{+}\rangle \left\langle \tilde{\Psi}_{+}\right\vert +\frac{%
\alpha }{7}\tilde{\sigma}_{+}+\frac{5-\alpha }{7}\tilde{\sigma}_{-}\right)
\right.
\end{equation}%
The evolution of $\sigma _{\alpha }$ in the presence of flipping noise and
its partial transpose can be found in a similar fashion as calculated for
equation (5). It is seen that the eigenvalues for each case remains
unchanged. Therefore, the local operation does not change the behavior of
entanglement in the presence of flipping noise.

\section{Conclusions}

Entanglement dynamics of a qutrit-qutrit system under the influence of
global, local and multilocal flipping noise is investigated. In order to
quantify the entanglement of the system, the negativity and realignment
criterion are used. It is shown that the entanglement sudden death and
distibility sudden death can be avoided in the presence of phase flip, trit
flip and trit phase flip environments. It is shown that certain free
entangled distillable qutrit states become bound entangled or separable i.e.
convert into non-distillable states under different flipping noises. It is
also seen that local operations do not have any effect on the entanglement
dynamics of the system. Further more, no ESD and DSD is seen for the case of
trit flip channel.\newline

{\huge Figures captions}\newline
\textbf{Figure 1}. (Color online). The negativity and realignment criterion
are plotted as a function of decoherence parameter $p$ for local noise
introduced through phase flip, trit flip and trit phase flip channels.%
\newline
\textbf{Figure 2}. (Color online). The negativity and realignment criterion
are plotted as a function of decoherence parameter $p$ for multi-local noise
introduced through phase flip, trit flip and trit phase flip channels.%
\newline
\textbf{Figure 3}. (Color online). The negativity and realignment criterion
are plotted as a function of decoherence parameter $p$ for global noise
introduced through phase flip, trit flip and trit phase flip channels.%
\newline
\textbf{Figure 4}. (Color online). The negativity is plotted as a function
of decoherence parameter $p$ and parameter $\alpha $ for multi-local and
global noises introduced through phase flip, trit flip and trit phase flip
channels.\newline
\newpage

\begin{figure}[tbp]
\begin{center}
\vspace{-2cm} \includegraphics[scale=0.6]{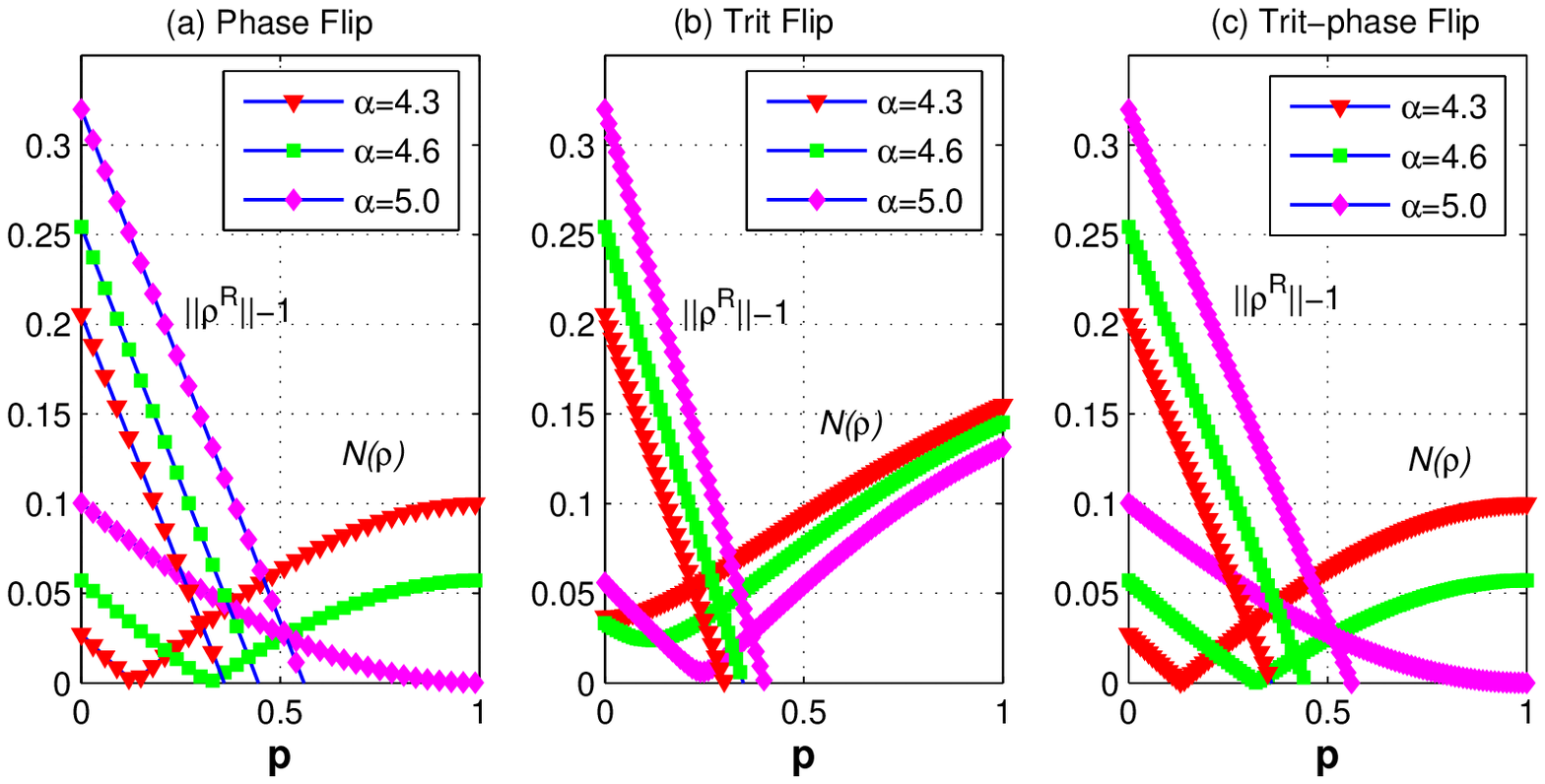} \\[0pt]
\end{center}
\caption{(Color online). The negativity and realignment criterion are
plotted as a function of decoherence parameter $p$ for local noise
introduced through phase flip, trit flip and trit phase flip channels.}
\end{figure}
\begin{figure}[tbp]
\begin{center}
\vspace{-2cm} \includegraphics[scale=0.6]{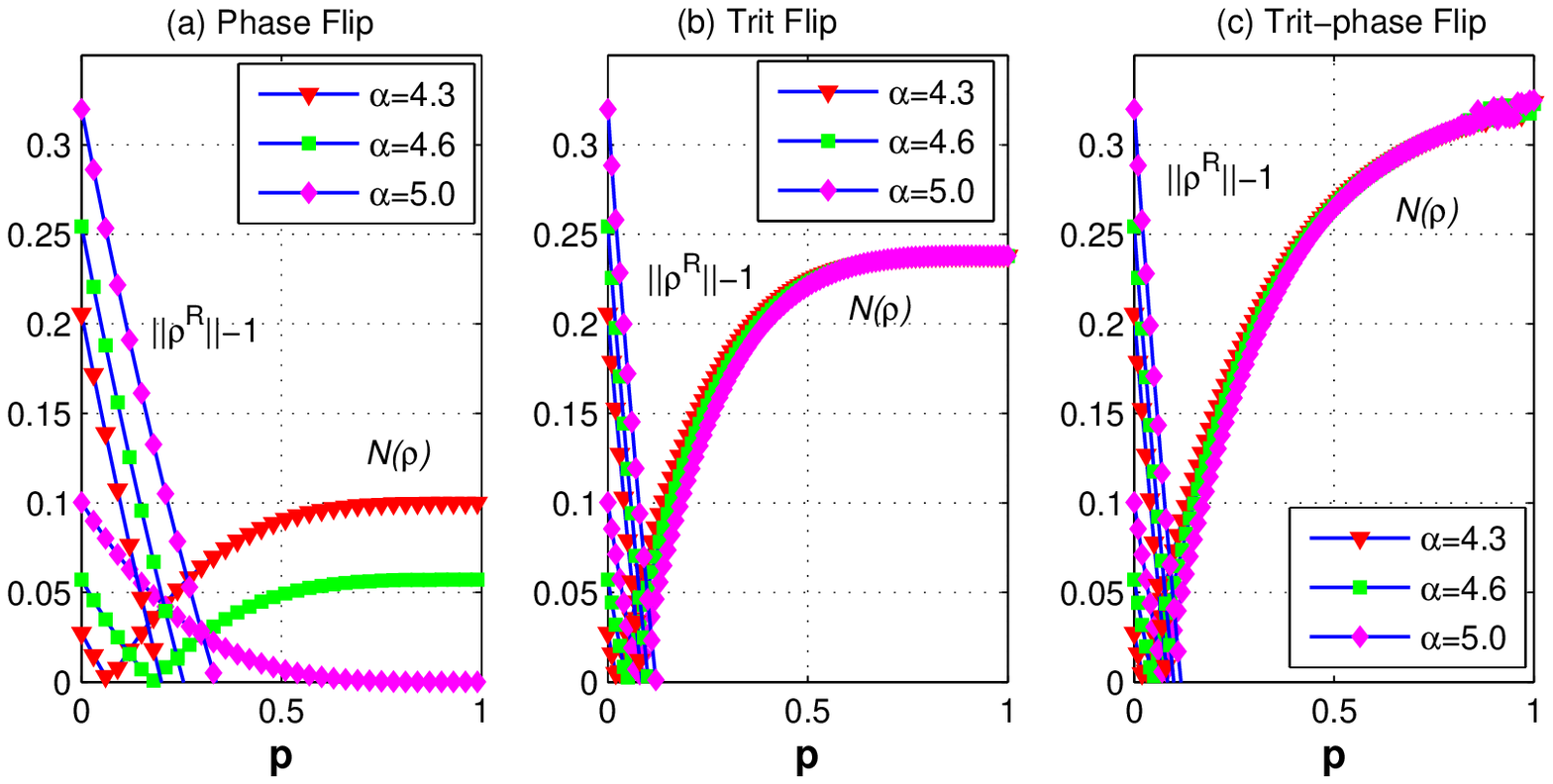} \\[0pt]
\end{center}
\caption{(Color online). The negativity and realignment criterion are
plotted as a function of decoherence parameter $p$ for multi-local noise
introduced through phase flip, trit flip and trit phase flip channels.}
\end{figure}
\begin{figure}[tbp]
\begin{center}
\vspace{-2cm} \includegraphics[scale=0.6]{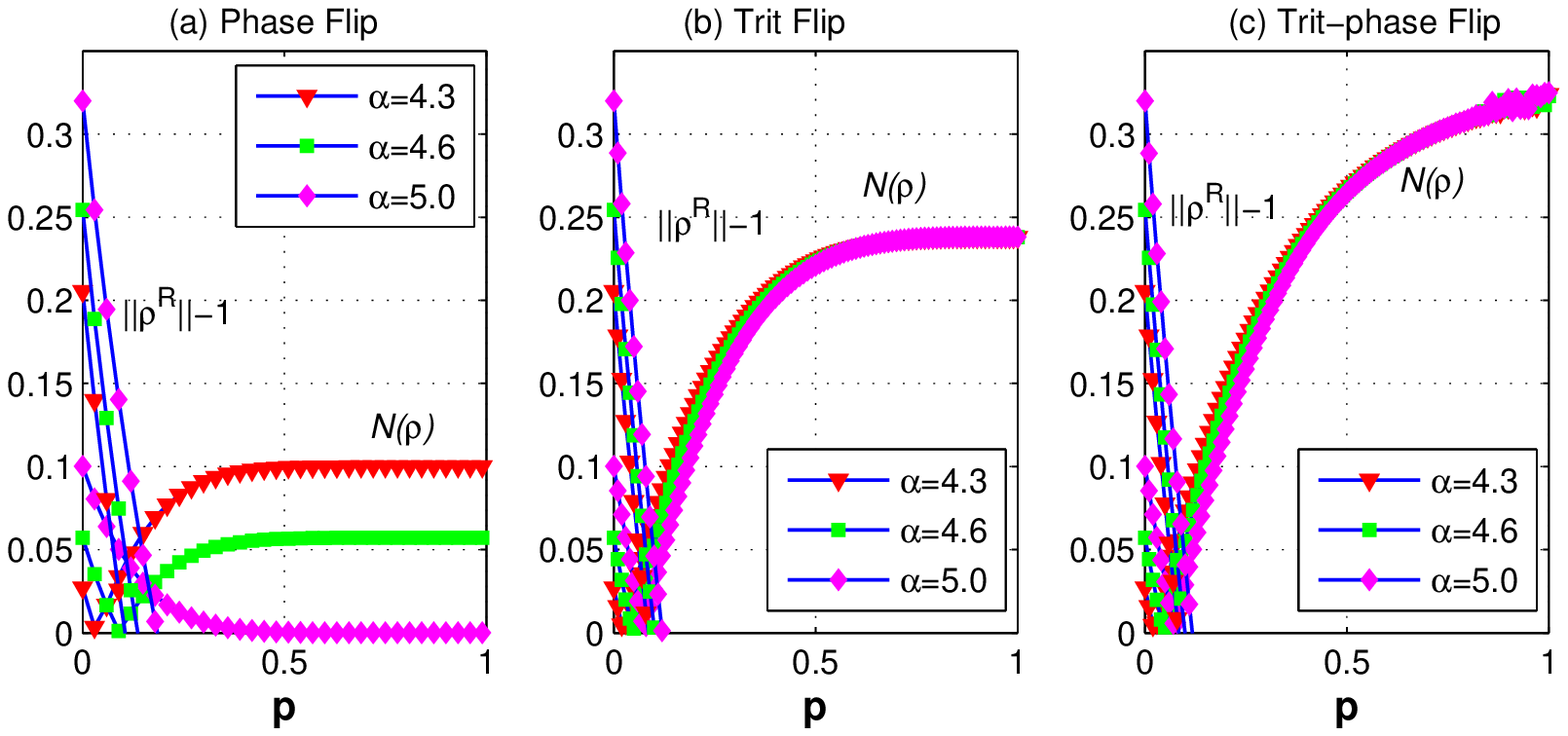} \\[0pt]
\end{center}
\caption{(Color online). The negativity and realignment criterion are
plotted as a function of decoherence parameter $p$ for global noise
introduced through phase flip, trit flip and trit phase flip channels.}
\end{figure}
\begin{figure}[tbp]
\begin{center}
\vspace{-2cm} \includegraphics[scale=0.6]{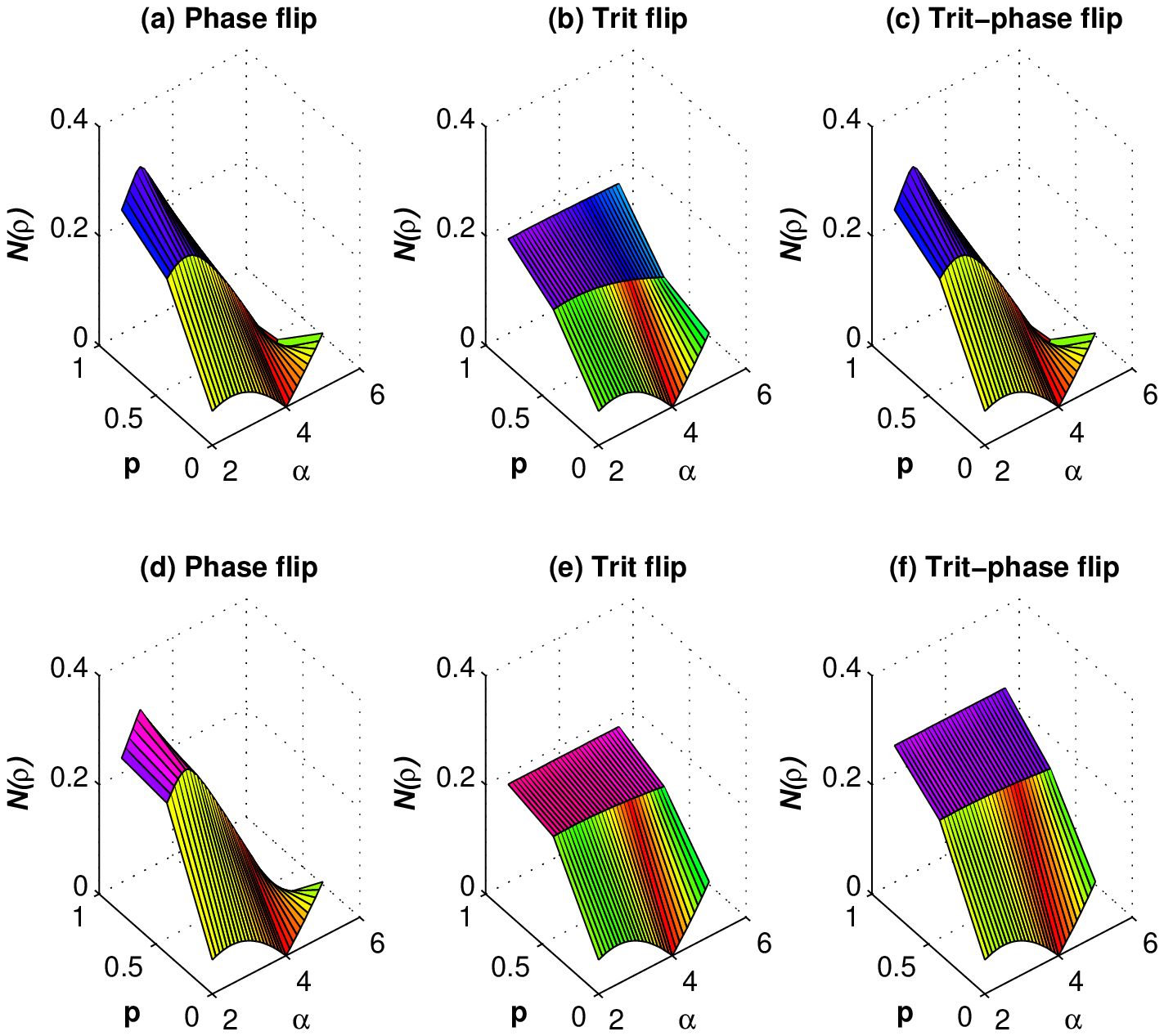} \\[0pt]
\end{center}
\caption{(Color online). The negativity is plotted as a function of
decoherence parameter $p$ and parameter $\protect\alpha $ for multi-local
and global noises introduced through phase flip, trit flip and trit phase
flip channels.}
\end{figure}


\begin{thebibliography}{99}
\bibitem{NieMA} Peres, A., Terno, D.R.: Rev. Mod. Phys. \textbf{76}, 93-123
(2004)

\bibitem{DB} Boschi, D., Branca, S., De Martini, F., Hardy, L., Popescu, S.:
Phys. Rev. Lett. \textbf{80}, 1121-1125 (1998)

\bibitem{DB2} Bouwmeester, D., Ekert, A., Zeilinger, A.: The Physics of
Quantum Information (Springer-Verlag, Berlin), (2000)

\bibitem{QEC} Preskill, J.: Proc. Roy. Soc. Lond. A \textbf{454,} 385-410
(1998)

\bibitem{KD} Ekert, A.: Phys. Rev. Lett. \textbf{67,} 661-663 (1991)

\bibitem{QC} Grover, L.K.: Phys. Rev. Lett. \textbf{79,} 325-328 (1997)

\bibitem{ESD1} Yonac, M., et al.: J. Phys. B \textbf{39,} S621-S625 (2006)

\bibitem{ESD2} Jakobczyk, L., Jamroz, A.: Phys. Lett. A \textbf{333,} 35-45
(2004)

\bibitem{ESD3} Ann, K., Jaeger, G.: Phys. Rev. A \textbf{76,} 044101 (2007)

\bibitem{ESD4} Jaeger, G., Ann, K.: J. Mod. Opt. \textbf{54,} 2327-2338
(2007)

\bibitem{HG} Yu, T., Eberly, J.H.: Phys. Rev. B \textbf{68,} 165322 (2003)

\bibitem{HB} Yu, T., Eberly, J.H.: Phys. Rev. Lett. \textbf{97,} 140403
(2007)

\bibitem{13} M.O. Terra Cunha, New J. Phys. 9 (2007) 237.

\bibitem{14} M.P. Almeida et al., Science 316 (2007) 579.

\bibitem{15} A. Salles, F. de Melo, M.P. Almeida, M. Hor-Meyll, S.P.
Walborn, P.H. Souto Ribeiro, L. Davidovich, Phys. Rev. A 78 022322 (2008)

\bibitem{16} J. Laurat, K.S. Choi, H. Deng, C.W. Chou, H.J. Kimble, Phys.
Rev. Lett. 99 180504 (2007)

\bibitem{PR} Peres, A.: Phys. Rev. Lett. \textbf{77,} 1413-1415 (1996)

\bibitem{HD} Horodecki, M., Horodecki, P., Horodecki, R.: Phys. Lett. A
\textbf{223,} 1-8 (1996)

\bibitem{Ann} Ann, K., Jeager, G.: Phys. Lett. A \textbf{372,} 579-583 (2008)

\bibitem{20} R. Horodecki, P. Horodecki, M. Horodecki, and K. Horodecki,
Rev. Mod. Phys. 81, 865 (2009)

\bibitem{21} M. Horodecki, P. Horodecki, and R. Horodecki, Phys. Rev. Lett.
80, 5239 (1998)

\bibitem{22} W. Song, L. Chen, and S. L. Zhu, Phys. Rev. A 80, 012331 (2009)

\bibitem{23} Ali M 2009 (Preprint arXiv:quant-ph/0911.0767)

\bibitem{24} Clarisse L Ph.D. thesis University of York, England, e-print
arXiv:quant-ph/0612072

\bibitem{25} Chen K and Wu L A 2003 Quantum Inf. Comput. 3 193-202; Rudolph
O, Quant. Info. Proc. 4 219-239 (2005)

\bibitem{26} G. Vidal and R. F. Werner, Phys. Rev. A 65, 032314 (2002)

\bibitem{28} Horodecki P, Horodecki M and Horodecki R, Phys. Rev. Lett. 82
1056--1059 (1999)

\bibitem{NC} Nielson, M.A., Chuang, I.L.: Quantum Computation and Quantum
Information (Cambridge: Cambridge University Press), (2000)\newpage
\end{thebibliography}
\end{document}